# A robust broadband fat suppressing phaser T$_2$ preparation module for cardiac magnetic resonance imaging at 3T

Lionel Arn[1], Ruud B van Heeswijk[1], Matthias Stuber[1,2], Jessica AM Bastiaansen[1]

*Submitted to Magnetic Resonance in Medicine*

[1]Department of Diagnostic and Interventional Radiology, Lausanne University Hospital and University of Lausanne, Lausanne, Switzerland

[2]Center for Biomedical Imaging, Lausanne, Switzerland

**To whom correspondence should be addressed:** Jessica AM Bastiaansen, Department of Diagnostic and Interventional Radiology, University Hospital Lausanne (CHUV), Rue de Bugnon 46, BH 08.074, 1011 Lausanne, Switzerland, Phone: +41-21-3147516, Email: jbastiaansen.mri@gmail.com, Twitter: @jessica_b_

**Category:** Note (max 2800 words, 5 figures including tables)

**Word count manuscript (body text):** 2717

**Figure count: 5**

**Table count: 0**

**Supporting figure count: 0**

## ABSTRACT (229 WORDS)

**Purpose**: Designing a new $T_2$ preparation (T2-Prep) module in order to simultaneously provide robust fat suppression and efficient $T_2$ preparation without requiring an additional fat suppression module for $T_2$-weighted imaging at 3T.

**Methods**: The tip-down RF pulse of an adiabatic $T_2$ preparation ($T_2$-Prep) module was replaced by a custom-designed RF excitation pulse that induces a phase difference between water and fat, resulting in a simultaneous $T_2$ preparation of water signals and the suppression of fat signals at the end of the module (now called a phaser adiabatic T2-Prep). Using numerical simulations, in vitro and in vivo ECG-triggered navigator gated acquisitions of the human heart, the blood, myocardium and fat signal-to-noise ratio and right coronary artery (RCA) vessel sharpness using this approach were compared against previously published conventional adiabatic $T_2$-Prep approaches

**Results**: Numerical simulations predicted an increased fat suppression bandwidth and decreased sensitivity against transmit magnetic field inhomogeneities using the proposed approach, while preserving the water $T_2$ preparation capabilities. This was confirmed by the tissue signals acquired on the phantom and the in vivo MRA, which show similar blood and myocardium SNR and CNR and significantly reduced fat SNR compared to the other methods tested. As a result, the RCA conspicuity was significantly increased and the motion artifacts were visually decreased.

**Conclusion**: A novel fat-suppressing $T_2$-preparation method was developed and implemented that demonstrated robust fat suppression and increased vessel sharpness compared with conventional techniques, while preserving its $T_2$ preparation capabilities.

## INTRODUCTION

In cardiac MRI, noncontrast-enhanced sequences often lack the required contrast to distinguish blood from myocardium. $T_2$ preparation ($T_2$-prep) modules (1) allow to increase this contrast by making use of the difference in $T_2$ relaxation time between these two tissue types (2,3), while they are also used for robust myocardial $T_2$ mapping at both 1.5T and 3T (4,5). Meanwhile, unwanted signal from fat can compromise the delineation and anatomical visualization of the coronary vessels and thus decrease the image quality in coronary MR angiography (MRA) (6). Unsuppressed fat signal may also lead to water-fat signal cancelation and can accentuate artefacts.

To suppress unwanted lipid signal, navigator-gated MRA sequences usually include additional fat signal suppressing approaches, such as CHEmically Selective Saturation (CHESS) (7), water-excitation (8–10), or frequency-selective inversion RF pulses (11,12). Higher magnetic field strengths however may complicate conventional fat saturation and $T_2$ preparation techniques, which are sensitive to $B_0$ and $B_1$ field inhomogeneities. The use of adiabatic RF excitation pulses addressed the sensitivity to $B_1$ inhomogeneities (13–16). A previous study in the thigh (15) showed that introducing a specific delay before the tip-up pulse of an adiabatic $T_2$-preparation module allowed the simultaneous saturation of spins resonating at a target frequency. However, the fat suppression bandwidth was inherently fixed and may not be sufficient to cope with $B_0$ inhomogeneities around the heart. Another study (17) demonstrated that decreasing the bandwidth of the tip-down RF excitation pulse resulted in an attenuation of off-resonance signals such as fat. However, because the off-resonance signal suppression was also dependent on a precise RF excitation angle, this method remained sensitive to $B_1$ inhomogeneities and still required the use of an additional fat saturation module. Moreover, the increase in RF excitation angle of both the tip-down and tip-up RF excitation pulse, which was required to negate inversion recovery of the fat signal in between the $T_2$-Prep and the start of the acquisition, reduced the efficiency of the $T_2$ preparation of on-resonance water (17).

In the current study, the aforementioned challenges of robustness to $B_0$ and $B_1$ inhomogeneities at 3T while preserving $T_2$ contrast and simultaneously suppressing fat are addressed by replacing the tip-down RF excitation pulse by a custom-designed pulse (Fig. 1A). The concept is to rotate both on- and off-resonance spins into the transverse plane while introducing a specific phase difference ($\Delta\varphi$) between the two populations of spins at the $T_2$ preparation echo time (Fig. 1B). Then, the non-selective tip-up pulse rotates the on-resonance magnetizations back along the longitudinal axis but because of $\Delta\varphi$, the off-resonance spins are rotated into the transverse plane and are spoiled by subsequent spoiler gradients (Fig. 1C). $\Delta\varphi$ can be adjusted to obtain the desired longitudinal components of the off-resonance magnetizations at the end of the module without altering the $T_2$ preparation of on-resonance spins and without increasing the RF power. This proposed fat-suppressing $T_2$ preparation technique, from here on referred to as the phaser adiabatic $T_2$-prep (PA-$T_2$-Prep), effectively rotates the fat magnetization with each applied RF excitation pulse, which increases the robustness to $B_1$ inhomogeneities.

The goal of this study was to implement and optimize the PA-$T_2$-Prep approach and to compare versus routinely used techniques. To this end, the efficacy of this approach was assessed by numerical simulations where the sensitivity of this technique to $B_0$ and $B_1$ inhomogeneities was quantified. The results from the simulations were validated in phantom and electrocardiogram (ECG)-triggered navigator-gated acquisitions of the human heart in healthy volunteers.

## METHODS

### *Tip-down RF pulse design*

The first RF pulse of the $T_2$-Prep was designed to simultaneously rotate the fat and water magnetization into the transverse plane with a total bandwidth comprising both water and fat while inducing a phase $\Delta\varphi$ between the two populations (Fig. 1D). The Fourier transform can be utilized to design small flip angle RF pulses. Hence, the pulse shape can be approximated as the inverse Fourier transform of two rectangular functions, each representing the desired transverse magnetization of water and fat, with bandwidth BW/2 shifted by ± BW/4 with a phase of ±$\Delta\varphi$/2. The pulse shape is thus proportional to the following description (Fig. 1E):

$$\frac{1}{2} sinc\left(\frac{2\pi\, BW/2}{2\pi}t\right)\left(e^{i\frac{\Delta\varphi}{2}}\,e^{-i\,2\pi\,t\,BW/4}\; +\; e^{-i\frac{\Delta\varphi}{2}}\,e^{i\,2\pi\,t\,BW/4}\right)$$

$$= \; sinc\left(\frac{BW}{2}t\right)cos\left(\pi\frac{BW}{2}t - \Delta\varphi/2\right) \qquad\qquad (1)$$

where *sinc* is the normalized cardinal sine function and t is time. The pulse duration PD was set to 8180 µs, the maximum duration allowed by the software. The pulse bandwidth was chosen in order to have a fat suppression bandwidth close to 700 Hz (total bandwidth 1400 Hz) while having zero-crossings at the beginning and the end of the pulse, which is satisfied if $BW = \frac{2k}{PD}, k \in \mathbb{Z}^*$. BW was set to 1467 Hz (i.e. k=3), the RF excitation angle was kept to 90° and the pulse off-resonance frequency ($f_0$) was set to -100 Hz, in between water (0 Hz) and fat (-407 Hz). The optimal $\Delta\varphi$ for water-fat phase separation, that facilitates fat suppression, which was adjustable on the user interface of the scanner, was subsequently determined empirically on a phantom such as to minimize the fat signal (see Methods - Phantom study).

To compensate for phase gain due to $f_0$ and $\Delta\varphi$, the initial phase of the phaser tip-down pulse was set to $-\frac{1}{2}(PD\ 2\ \pi\ f_0\ +\ \Delta\varphi)$.

## Numerical simulations

The effects of the conventional adiabatic T2-prep (CA-$T_2$-Prep) (14) with CHESS fat saturation (+ FS) and the proposed PA-$T_2$-Prep on the longitudinal magnetization were quantified using numerical simulations performed in MATLAB (The Mathworks, Natick, MA, USA). Both the tip-down and tip-up RF excitation pulses were modelled as Hamming-windowed sinc functions of 800 µs of opposite polarity for the CA-$T_2$-Prep + FS. For the PA-$T_2$-Prep, the tip-down pulse as described above was used while the tip-up pulse was identical to that of the CA-$T_2$-Prep. For both methods, the two hyperbolic secant refocusing adiabatic pulses were modelled as ideal 180° rotations. The CHESS pulse was modelled as implemented on the scanner, a 5120 µs, -407 Hz Gaussian pulse with RF excitation angle = 100°. The simulations were performed by Euler integration of the Bloch equations using steps of 1 µs. The $T_1$ and $T_2$ relaxation times were set to those of fat at 200 ms and 50 ms, respectively, while the off-resonance frequency was varied from -1000 Hz to 1000 Hz, and the $B_1$ amplitude ranged from 0 to 200% to account for $B_0$ and $B_1$ magnetic field inhomogeneities. The peak frequency of fat was assumed to resonate at -407 Hz off resonance and a magnetization was considered suppressed if its longitudinal component at the end of the $T_2$-prep was reduced to between ± 10% of the starting magnetization $M_0$.

## Phantom study

The optimal water-fat specific phase difference $\Delta\varphi$ was empirically searched using a three-compartment cylindrical phantom containing mixed solutions of agar and $NiCl_2$ (Sigma-Aldrich,

St. Louis, MO) (18), or baby oil (Johnson and Johnson, New Brunswick, NJ), which mimics the magnetic relaxation properties of blood ($T_1$=1463 ± 24 ms, $T_2$=221 ± 15 ms), muscle ($T_1$=1072 ± 8 ms, $T_2$=31 ± 0.4 ms) , and fat ($T_1$=211 ±1 ms, $T_2$=42±0.3 ms). The phantom acquisition was performed on a clinical 3T scanner (MAGNETOM Prisma[fit], Siemens Healthcare, Erlangen, Germany) using a 18 channel spine and 16 channel chest RF coil with a segmented 3D Cartesian GRE sequence with a field of view (FOV) of 160 × 96 × 95 $mm^3$ and resolution of 1 x 1 x 5 $mm^3$. After a 40 ms $T_2$-prep with 10 ms hyberbolic secant adiabatic refocusing pulses (13,5), 25 k-space lines were acquired using centric ordering. TR = 5 ms, TE = 2.5 ms, RF excitation angle = 15°, bandwidth = 501 Hz/pixel, time between $T_2$-preps = 1 s. $\Delta\varphi$ was varied between 90° to 270° by steps of 5° (randomized acquisition order) and the resulting average signal from each of the 3 compartments was quantified. The $\Delta\varphi$ yielding the lowest fat signal in this experiment was used for the rest of the study.

Next, the mean signals in all phantom compartments were quantified after using five different $T_2$ preparation modules: no $T_2$-Prep, CA-$T_2$-Prep, CA-$T_2$-Prep + FS, the proposed PA-$T_2$-prep and the water-selective adiabatic $T_2$-Prep with fat saturation (WSA-$T_2$-Prep + FS) (17). The $B_0$ and $B_1$ fields were intentionally not shimmed to test the sensitivity of each technique to inhomogeneities. A one-hour long scan with 100 averages was acquired prior to the following acquisitions to reach thermal steady state for true noise quantification in scan-rescan acquisitions. Scan-rescan acquisitions were performed for quantification of blood, myocardium and fat compartment mean signal and noise using the acquisition-subtraction method (19). A compartment specific SNR was computed relative to the no $T_2$-Prep case to emphasize compartments specific differences across $T_2$-preps. A $B_0$ map with a bandwidth of ± 1000 Hz was computed from two acquisitions with the same acquisition parameters as described previously, but with continuous excitation, without $T_2$-Prep and with a TE of 2.5 ms and 3.0 ms.

Scaling in $B_0$ maps was cropped to $\pm 500$ Hz. Additionally, a $B_1$ map from two acquisitions with RF excitation angles of 60° and 120° was computed, with a TR of 10 s using a double-angle method (20).

### *In Vivo study*

Whole-heart free-breathing ECG-triggered navigator-gated scans were acquired in healthy volunteers (N=6, age=27±4y). The volunteers gave written and informed consent. The study was carried out according to the institutional rules. In order to keep the scan time below one hour per participant, only the CA-$T_2$-Prep, CA-$T_2$-Prep + FS and PA-$T_2$-Prep were tested. The same imaging sequence plus parameters were used as described in the phantom study, except for a FOV of 230x368x86.4 $mm^3$ and an isotropic pixel size of 1.2 mm. After acquisition, the images were reformatted using Soap-Bubble, a semi-automated reformatting and vessel tracking software package (21). The right coronary artery (RCA) vessel sharpness and blood, myocardium, chest fat and epicardial SNR were quantified in chest fat, epicardial fat, blood and myocardium. SNR was approximated by dividing the signal averages from regions of interests drawn in respective tissues by the standard deviation of the background noise. Differences between the CA-$T_2$-Prep and PA-$T_2$-Prep as well as between the CA-$T_2$-Prep + FS and PA-$T_2$-Prep were tested via a Student's t-test for paired data using the Bonferroni correction with a p-value < 0.05 considered statistically significant. Additionally, the specific absorption rate (SAR) was recorded from the system console.

## RESULTS

### Numerical simulation

The simulation of the PA-$T_2$-Prep (Fig. 2A) predicted an increased robustness in fat suppression against both $B_0$ and $B_1$ field inhomogeneities in comparison to the CA-$T_2$-Prep + FS (Fig. 2B). For a tissue with an off-resonance frequency of -407 Hz, the magnetization was suppressed if $B_1$ was between 84% and 111% for the CA-$T_2$-Prep + FS, versus 75% and 130% for the PA-$T_2$-Prep. At a $B_1$ amplitude of 100%, the fat suppression bandwidth was 162 Hz for the CA-$T_2$-prep + FS and 627 Hz for the PA-$T_2$-Prep.

### Phantom study

The minimum average fat signal in the phantom occurred at a $\Delta\varphi$ of 110° (Fig. 3A). This parameter was kept for the remainder of this study. The SNR of the blood, muscle, and fat compartment using either the CA-$T_2$-Prep, CA-$T_2$-Prep + FS, WSA-$T_2$-Prep + FS and PA-$T_2$-Prep relative to using no $T_2$-Prep show differences in fat suppression and $T_2$-weighting (Fig. 3B). Compared to the CA-$T_2$-Prep, CA-$T_2$-Prep + FS and WSA-$T_2$-Prep+FS, the PA-$T_2$-Prep reduced the lipid signal by 93.2%, 70.0% and 66.1% respectively. The blood and myocardium compartment SNR in PA-$T_2$-Prep images were respectively 0.7% and 0.9% lower than in the CA-$T_2$-Prep, and 3.4% and 26.9% lower than in the WSA-$T_2$-Prep+FS.

Fat suppression using the PA-$T_2$-Prep was improved in regions where field inhomogeneities hindered the capabilities of other methods (Fig. 4, red arrows). In contrast to the CA-$T_2$-Prep + FS and PA-$T_2$-Prep, the WSA-$T_2$-Prep reduced blood and myocardium signals in regions where the $B_0$ imperfections increased the precession frequency by more than 200 Hz (Fig. 4, green arrows).

*In Vivo study*

Compared to the CA-$T_2$-Prep+FS, the PA-$T_2$-Prep reduced chest and epicardial fat SNR by 53% (p=0.03) and 47% (p=0.02) respectively in volunteers. No significant differences between blood SNR (p=0.29), myocardium SNR (p=0.34) and blood-myocardium CNR (p=0.30) were observed. The more homogeneous suppression of fat signal increased vessel sharpness by 24% (p=0.01). Compared to the non-fat suppressing CA-$T_2$-Prep, the PA-$T_2$-Prep reduced chest and epicardial fat SNR by 79% (p=0.001) and 74% (p=0.001) respectively and increased vessel sharpness by 36% (p=0.005). No significant differences between blood SNR (p=0.66), myocardium SNR (p=0.90) and blood-myocardium CNR (p=0.35) were observed. The PA-$T_2$-Prep visually reduced the amount of motion artefacts in the image (Fig. 5). The SAR increased by 1% using the PA-$T_2$-Prep compared to the two other methods.

## DISCUSSION

As demonstrated by the Bloch equation simulations, the PA-$T_2$-Prep not only increases the fat saturation bandwidth, but also avoids water signal attenuation when its resonance frequency is shifted due to field inhomogeneities. Additionally, the multiple rotations of the fat magnetization during the PA-$T_2$-Prep decrease the fat suppression dependence on $B_1$ precision compared to the use of a single fat saturation pulse.

The WSA-T2-Prep (17) achieves precise fat suppression by increasing the RF excitation angle of both tip-down and tip-up pulses. However, this approach also affects the on-resonance magnetization and as a result, some loss in $T_2$ preparation can be expected. This effect was observed and confirmed in the phantom experiments. Because the PA-$T_2$-Prep rotates the magnetization 90°, the effectiveness of the $T_2$ preparation is conserved, as confirmed by the measured SNR in the phantom and in vivo experiments. In phantom experiments it was demonstrated that the water-fat phase separation $\Delta\varphi$ influences the final orientation of the fat magnetization after the T2-Prep (Fig. 1). This parameter can be adjusted to adapt the module to different pulse sequences. For example, a balanced steady-state free precession acquisition may require ramp-up pulses between the $T_2$-Prep and the acquisition, during which the fat signal may recover. In this case, $\Delta\varphi$ may be increased to compensate for this additional recovery. In the phantom experiments, the slight decrease of <1% (yet above the noise level) in blood and myocardium compartment signal using PA-$T_2$-Prep may be explained by off-resonance partial volume suppression. Another possible explanation for this loss could be the excitation profile ripples of the phaser tip-down pulse due to its finite duration, which are not compensated during the tip-up pulse. In any case, this effect is relatively small and can be neglected.

Similar to the CA-$T_2$-Prep with FS or the WSA-$T_2$-Prep, the new proposed method only suppresses the fat once before the acquisition. As such, the effectiveness of the fat suppression is ultimately limited by the pulse sequence that follows. In sequences using Cartesian sampling trajectories the proposed PA-$T_2$-Prep may remove the need for additional fat suppression modules in cardiac imaging and thus reduce SAR. However, it will be less suitable for acquisition schemes that do not allow for centric reordering of the phase encoding planes of k-space, such as radial acquisitions. Although water-selective excitation approaches may be more suitable for fat suppression in radial whole-heart MRI at 3T (22–24), acquisitions requiring $T_2$-preparation modules may still benefit from the proposed approach as it can offer an additional, tunable range of fat suppression. More recently $T_2$-Prep approaches designed for coronary artery imaging at 1.5 included outer volume suppression strategies (25,26), with integrated spectral spatial sinc pulses for fat suppression (27). Their utility at 3T remains to be investigated considering SAR may be higher and bSSFP less performant at higher field strengths, therefore these methods were not compared in the current study. Another limitation includes the minimum duration of the $T_2$-prep module that the lengthy tip-down pulse imposes. Compared to the minimum duration of 21.7 ms for the CA-$T_2$-prep, the PA-$T_2$-prep requires at least 29.0 ms, which limits the range of available $T_2$ contrast, although this may be addressed by using a shorter adiabatic refocusing pulse pair. However, this is not a limitation for coronary MRA where longer $T_2$-Prep durations are typically used, but its effects on $T_2$ mapping may have to be investigated.

Nevertheless, the proposed fat-suppressing PA-$T_2$-prep may indeed benefit $T_2$ mapping techniques that utilize incrementing $T_2$-prep durations. This may reduce chemical shift artifacts that hinder cartilage delineation and quantification in human knees (28), and may obviate the use of additional fat suppression modules in cardiac $T_2$ mapping (4,5,29) or combined $T_1$ and $T_2$ mapping techniques (30,31).

## CONCLUSION

In this study, a novel $T_2$ preparation approach was developed that provides a large spectral bandwidth of fat suppression. As confirmed by simulations as well as in vitro and in vivo acquisitions, the technique is robust against $B_0$ and $B_1$ inhomogeneities, and preserves blood-myocardium SNR and CNR. As a result, the RCA vessel sharpness was increased compared with conventional approaches.

## ACKNOWLEDGEMENTS

JB received funding from the Swiss National Science Foundation (grant number PZ00P3_167871), the Emma Muschamp foundation, and the Swiss Heart foundation. RBvH received funding from the Swiss National Science Foundation (grant number 32003B_182615) and MS from the Swiss National Science Foundation (grant numbers 320030_173129 and 326030_150828).

# FIGURES

*Figure 1*

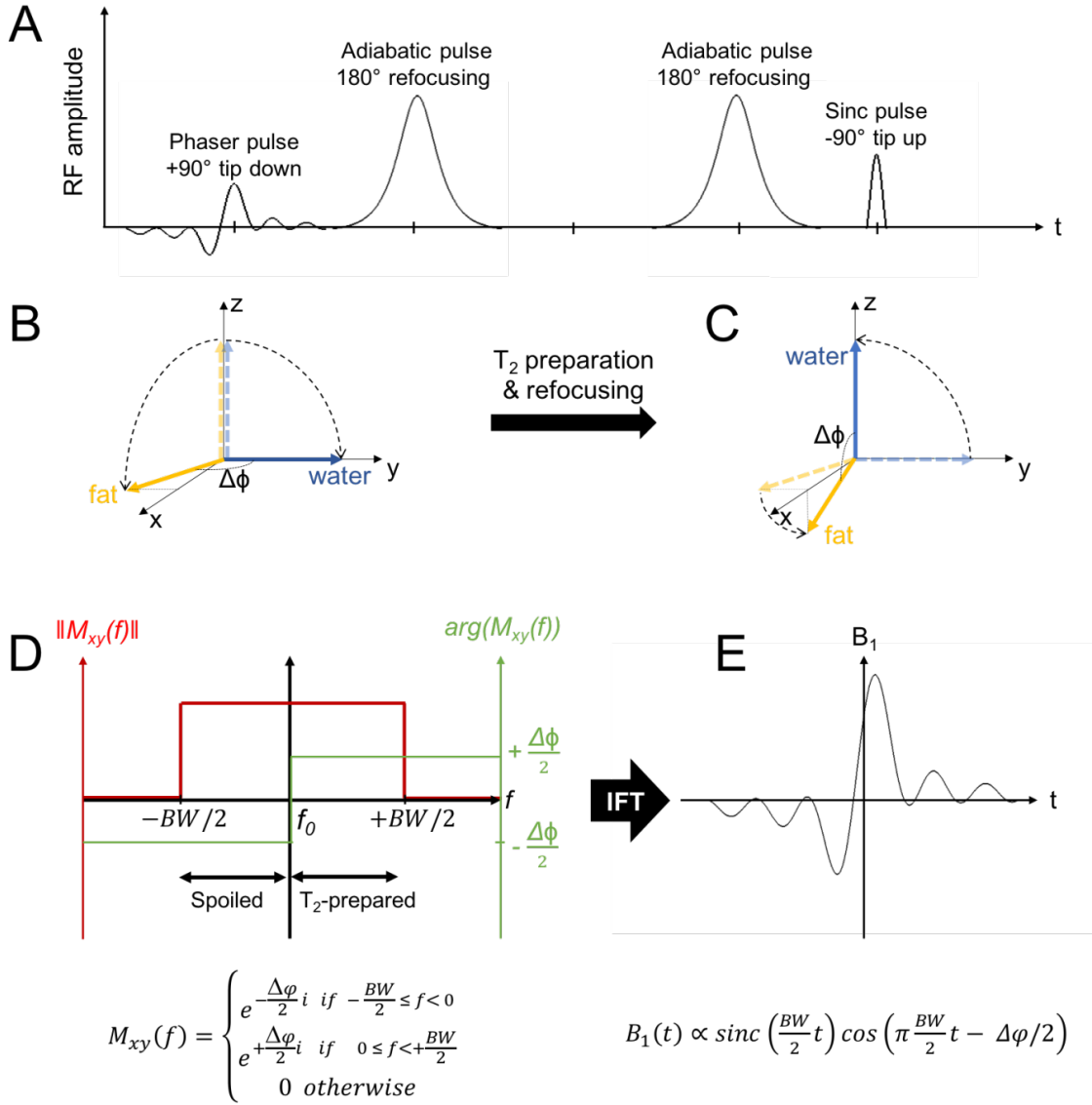

$$M_{xy}(f) = \begin{cases} e^{-\frac{\Delta\varphi}{2}i} & if \quad -\frac{BW}{2} \leq f < 0 \\ e^{+\frac{\Delta\varphi}{2}i} & if \quad 0 \leq f < +\frac{BW}{2} \\ 0 \quad otherwise \end{cases}$$

$$B_1(t) \propto sinc\left(\frac{BW}{2}t\right)cos\left(\pi\frac{BW}{2}t - \Delta\varphi/2\right)$$

**Diagram of the proposed phaser T2 preparation module and design of the phaser tip-down RF pulse**. **(A)** Overview of RF pulses used in the T2 preparation module. The module is followed by a spoiling gradient (not illustrated). **(B)** Excitation of the water (blue) and fat (yellow) magnetizations by the phaser tip-down pulse. The solid and dashed vectors represent the magnetization before and after the excitation respectively, and the black dashed trajectories represent their rotations. Both fat and water are rotated into the transverse plane, but the pulse introduces a phase difference $\Delta\varphi$ between the two. **(C)** Excitation of the water and fat magnetizations by broadband tip-up pulse. The $T_2$ prepared water magnetization is restored along the longitudinal axis while the fat is kept near the transverse plane because of $\Delta\varphi$. **(D)** Desired transverse magnetization fraction $\|M_{xy}(f)\|$ (red) and magnetization phase $arg(M_{xy}(f))$ (green) after the phaser tip-down RF excitation as function of spin off-resonance frequency. BW: RF excitation bandwidth, $\Delta\varphi$: phase difference between fat and water magnetization, $f_0$: RF excitation pulse frequency **(E)** $B_1$ amplitude of the phaser tip-down RF excitation pulse as a function of time, calculated as the inverse Fourier transform of **(D)**.

*Figure 2*

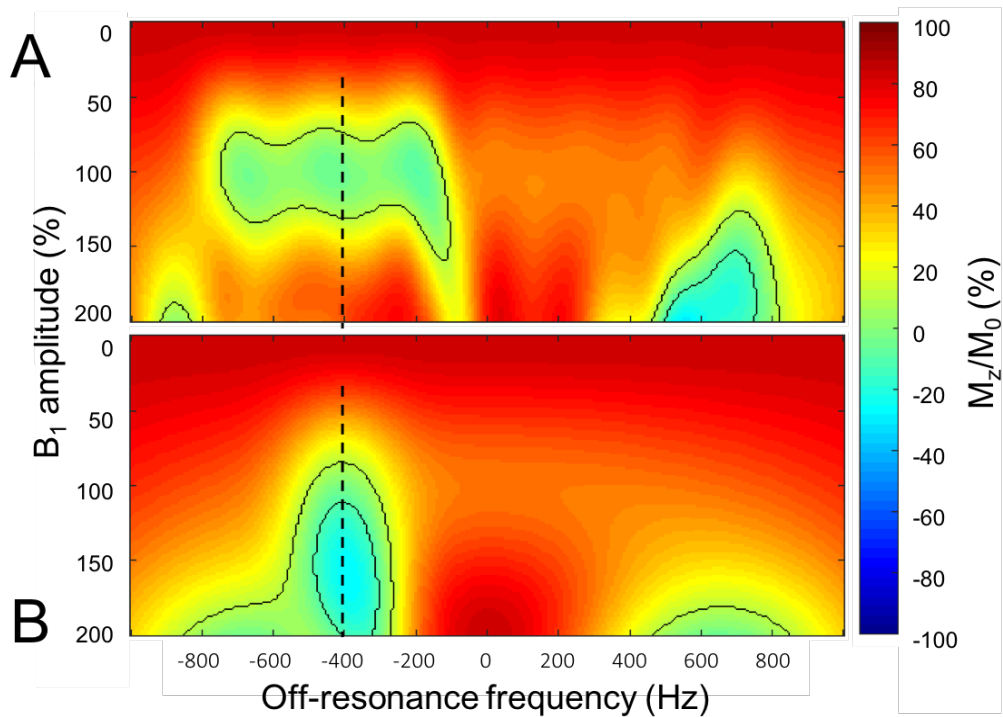

Numerical simulation of the performance of the phaser adiabatic $T_2$ prep (PA-$T_2$-Prep) and the conventional adiabatic $T_2$-prep with fat saturation (CA-$T_2$-Prep+FS). The longitudinal magnetization ($M_z$) as a fraction of the initial magnetization ($M_0$) is plotted as a function of $B_1$ strength and tissue off-resonance frequency after application of **(A)** the PA-$T_2$-prep and **(B)** CA-$T_2$-prep + FS. A $B_1$ value of 100% corresponds to the desired applied RF amplitude. The dashed line indicates the center frequency of fat (-407 Hz), and the outlined green regions correspond to a resulting longitudinal magnetization within ±10% of the initial magnetization.

*Figure 3*

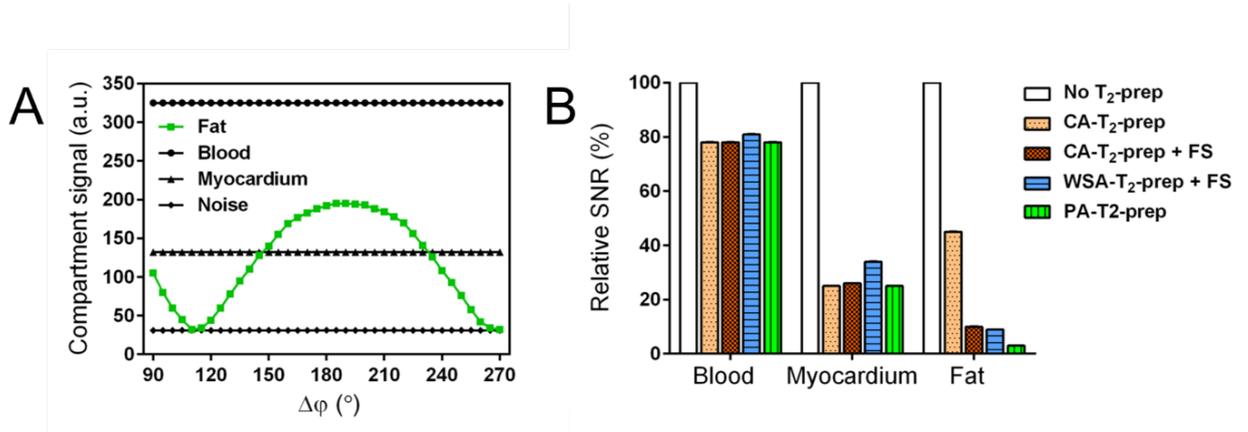

(A) Blood, myocardium and fat compartments as well as noise mean signal extracted from the images of the multi-compartment phantom. Local minima on the fat signal curve indicate optimal fat suppression for $\Delta\varphi=110°$ and $\Delta\varphi=270°$, where the intensity reaches the noise level. The blood and myocardium compartment signal is largely unaffected by the change of $\Delta\varphi$. (B) SNR of the blood, myocardium and fat compartments of the phantom using different $T_2$-Prep techniques relative to using no $T_2$-Prep. The blood and myocardium signals in the CA-$T_2$-Prep, CA-$T_2$-Prep+FS and PA-$T_2$-Prep were similar. However, the 120° RF excitation angle of WSA-$T_2$-Prep reduced the efficiency of the $T_2$ preparation, resulting in increased myocardium SNR. The fat signal was most reduced using the PA-$T_2$-Pprep.

*Figure 4*

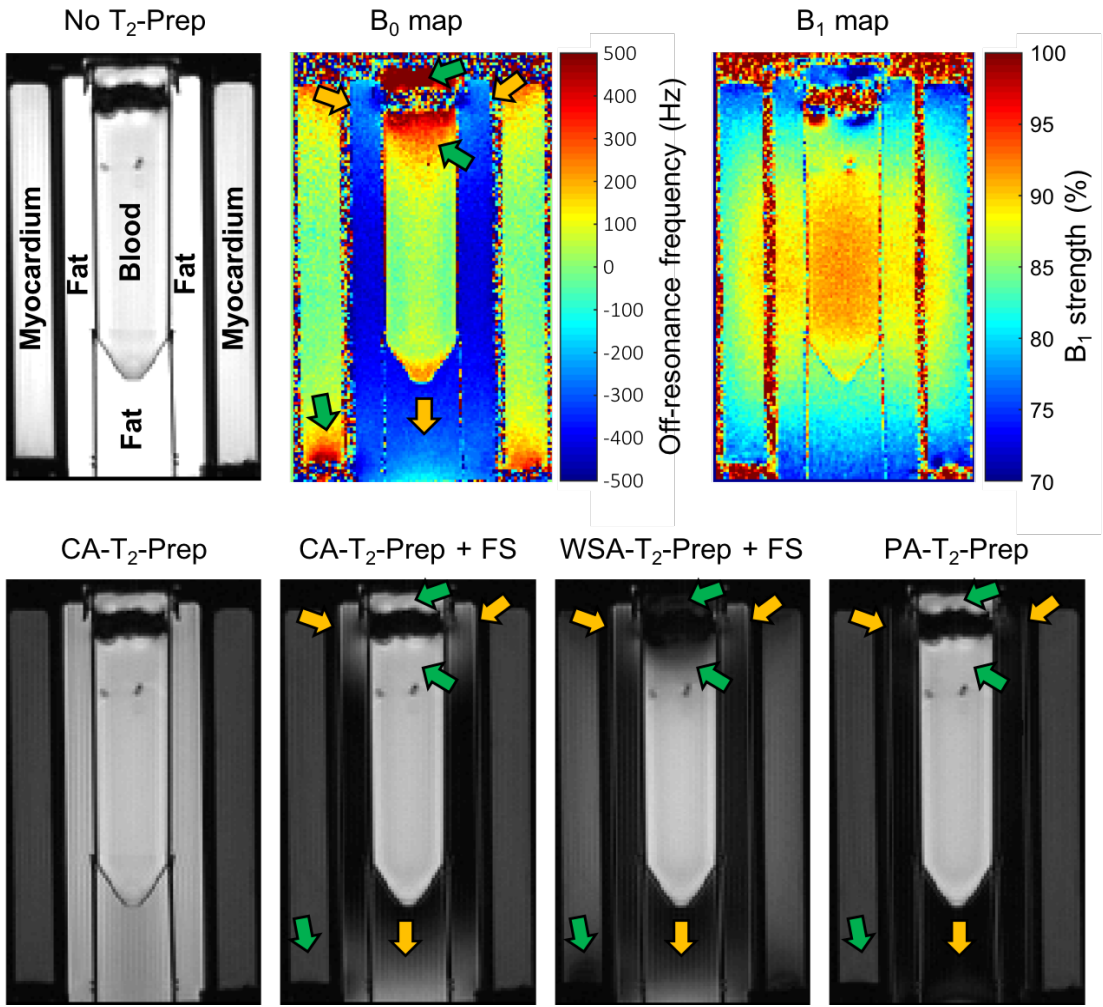

A comparison of the different T2 preparation methods in the multi-compartment phantom with corresponding $B_0$ and $B_1$ maps..  Images acquired with no $T_2$-Prep, the conventional adiabatic $T_2$-Prep (CA-$T_2$-Prep) without and with fat saturation (FS), the water-selective adiabatic $T_2$-Prep (WSA-$T_2$-Prep) and the phaser adiabatic $T_2$-Prep (PA-$T_2$-Prep). Orange arrows indicate locations of improved fat suppression using the PA-$T_2$-Prep compared to other fat suppression methods. Green arrows indicate positive off-resonance regions in which the water selective method attenuated the signal which was not observed using the PA-$T_2$-prep.

*Figure 5*

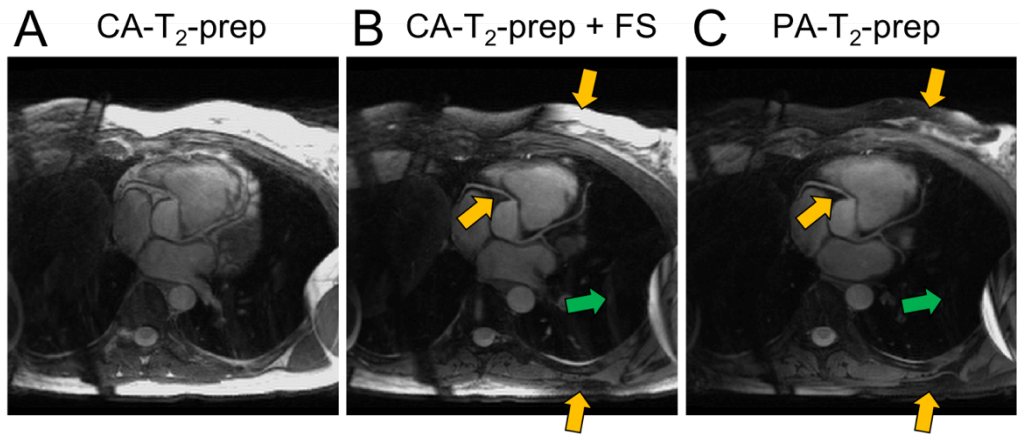

A comparison of the $T_2$ preparation modules in the human heart in vivo at 3T. ECG-triggered navigator-gated reformatted images of a healthy volunteer using the conventional adiabatic $T_2$-Prep without and with fat sat (**A,B**) and the phaser adiabatic $T_2$-Prep (PA-$T_2$-Prep) (**C**). Orange arrows indicate improved fat suppression using the PA-$T_2$-Prep compared to the conventional fat suppression (FS) using CHESS. Green arrows show a decrease of motion artefact when the lipid intense signal was further attenuated.

# REFERENCES

1. Brittain JH, Hu BS, Wright GA, Meyer CH, Macovski A, Nishimura DG. Coronary Angiography with Magnetization-Prepared T2 Contrast. Magnetic Resonance in Medicine 1995;33:689–696 doi: 10.1002/mrm.1910330515.

2. Shea SM, Deshpande VS, Chung Y-C, Li D. Three-dimensional true-FISP imaging of the coronary arteries: Improved contrast with T2-preparation. Journal of Magnetic Resonance Imaging 2002;15:597–602 doi: 10.1002/jmri.10106.

3. Botnar RM, Stuber M, Danias PG, Kissinger KV, Manning WJ. Improved coronary artery definition with T2-weighted, free-breathing, three-dimensional coronary MRA. Circulation 1999;99:3139–48.

4. Foltz WD, Al-Kwifi O, Sussman MS, Stainsby JA, Wright GA. Optimized spiral imaging for measurement of myocardial T2 relaxation. Magnetic Resonance in Medicine 2003;49:1089–1097 doi: 10.1002/mrm.10467.

5. van Heeswijk RB, Feliciano H, Bongard C, et al. Free-Breathing 3 T Magnetic Resonance T2-Mapping of the Heart. JACC: Cardiovascular Imaging 2012;5:1231–1239 doi: 10.1016/j.jcmg.2012.06.010.

6. Manning W J, Li W, Boyle N G, Edelman R R. Fat-suppressed breath-hold magnetic resonance coronary angiography. Circulation 1993;87:94–104 doi: 10.1161/01.CIR.87.1.94.

7. Haase A, Frahm J, Hanicke W, Matthaei D. H-1-Nmr Chemical-Shift Selective (Chess) Imaging. Phys Med Biol 1985;30:341–344 doi: 10.1088/0031-9155/30/4/008.

8. Meyer CH, Pauly JM, Macovski A, Nishimura DG. Simultaneous spatial and spectral selective excitation. Magn Reson Med 1990;15:287–304.

9. Schick F. Simultaneous highly selective MR water and fat imaging using a simple new type of spectral-spatial excitation. Magn Reson Med 1998;40:194–202.

10. Bastiaansen JAM, Stuber M. Flexible water excitation for fat-free MRI at 3T using lipid insensitive binomial off-resonant RF excitation (LIBRE) pulses. Magn Reson Med 2017 doi: 10.1002/mrm.26965.

11. van Elderen SGC, Versluis MJ, Westenberg JJM, et al. Right Coronary MR Angiography at 7 T: A Direct Quantitative and Qualitative Comparison with 3 T in Young Healthy Volunteers. Radiology 2010;257:254–259 doi: 10.1148/radiol.100615.

12. Bhat H, Yang Q, Zuehlsdorff S, Li K, Li D. Contrast-Enhanced Whole-Heart Coronary MRA at 3T Using Interleaved EPI. Invest Radiol 2010;45:458–464 doi: 10.1097/RLI.0b013e3181d8df32.

13. Hwang TL, van Zijl PC, Garwood M. Fast broadband inversion by adiabatic pulses. J. Magn. Reson. 1998;133:200–203 doi: 10.1006/jmre.1998.1441.

14. Nezafat R, Stuber M, Ouwerkerk R, Gharib AM, Desai MY, Pettigrew RI. B1-insensitive T2 preparation for improved coronary magnetic resonance angiography at 3 T. Magn Reson Med 2006;55:858–64 doi: 10.1002/mrm.20835.

15. Nezafat R, Ouwerkerk R, Derbyshire AJ, Stuber M, McVeigh ER. Spectrally selective B1-insensitive T2 magnetization preparation sequence. Magn Reson Med 2009;61:1326–1335 doi: 10.1002/mrm.21742.

16. Soleimanifard S, Schär M, Hays AG, Prince JL, Weiss RG, Stuber M. A Spatially-selective Implementation of the Adiabatic T2Prep Sequence for Magnetic Resonance Angiography of the Coronary Arteries. Magn Reson Med 2013;70:97–105 doi: 10.1002/mrm.24437.

17. Coristine AJ, van Heeswijk RB, Stuber M. Fat signal suppression for coronary MRA at 3T using a water-selective adiabatic T2 -preparation technique. Magnetic resonance in medicine 2014;72:763–9 doi: 10.1002/mrm.24961.

18. Kraft KA, Fatouros PP, Clarke GD, Kishore PR. An MRI phantom material for quantitative relaxometry. Magnetic resonance in medicine 1987;5:555–62.

19. Firbank MJ, Coulthard A, Harrison RM, Williams ED. A comparison of two methods for measuring the signal to noise ratio on MR images. Phys Med Biol 1999;44:N261-264 doi: 10.1088/0031-9155/44/12/403.

20. Cunningham CH, Pauly JM, Nayak KS. Saturated double-angle method for rapid B1+ mapping. Magn Reson Med 2006;55:1326–1333 doi: 10.1002/mrm.20896.

21. Etienne A, Botnar RM, Van Muiswinkel AM, Boesiger P, Manning WJ, Stuber M. "Soap-Bubble" visualization and quantitative analysis of 3D coronary magnetic resonance angiograms. Magn Reson Med 2002;48:658–66 doi: 10.1002/mrm.10253.

22. Pang J, Sharif B, Fan Z, et al. ECG and navigator-free four-dimensional whole-heart coronary MRA for simultaneous visualization of cardiac anatomy and function. Magn Reson Med 2014;72:1208–17 doi: 10.1002/mrm.25450.

23. Bastiaansen JAM, van Heeswijk RB, Stuber M, Piccini D. Noncontrast free-breathing respiratory self-navigated coronary artery cardiovascular magnetic resonance angiography at 3 T using lipid insensitive binomial off-resonant excitation (LIBRE). Journal of Cardiovascular Magnetic Resonance 2019;21:38 doi: 10.1186/s12968-019-0543-6.

24. Bastiaansen JAM, Piccini D, Sopra LD, et al. Natively fat-suppressed 5D whole-heart MRI with a radial free-running fast-interrupted steady-state (FISS) sequence at 1.5T and 3T. Magnetic Resonance in Medicine 2020;83:45–55 doi: 10.1002/mrm.27942.

25. Colotti R, Omoumi P, van Heeswijk RB, Bastiaansen JAM. Simultaneous fat-free isotropic 3D anatomical imaging and T2 mapping of knee cartilage with lipid-insensitive binomial off-resonant RF excitation (LIBRE) pulses. J Magn Reson Imaging 2018 doi: 10.1002/jmri.26322.

26. Giri S, Chung Y-C, Merchant A, et al. T2 quantification for improved detection of myocardial edema. Journal of Cardiovascular Magnetic Resonance 2009;11:56 doi: 10.1186/1532-429X-11-56.

27. Kvernby S, Warntjes MJB, Haraldsson H, Carlhäll C-J, Engvall J, Ebbers T. Simultaneous three-dimensional myocardial T1 and T2 mapping in one breath hold with 3D-QALAS. Journal of Cardiovascular Magnetic Resonance 2014;16:102 doi: 10.1186/s12968-014-0102-0.

28. Hamilton JI, Jiang Y, Chen Y, et al. MR fingerprinting for rapid quantification of myocardial T1, T2, and proton spin density. Magnetic Resonance in Medicine 2017;77:1446–1458 doi: 10.1002/mrm.26216.